# The Global Landscape of Academic Guidelines for Generative AI and Large Language Models

**Junfeng Jiao [1]  Saleh Afroogh*[2]  Kevin Chen [3]  David Atkinson [4]  Amit Dhurandhar [5]**

1. Urban Information Lab, The University of Texas at Austin, Austin, TX 78712, United 2. States. jjiao@austin.utexas.edu
2. Urban Information Lab, The University of Texas at Austin, Austin, TX 78712, 4. United States. Saleh.afroogh@utexas.edu
3. Department of economics, The University of Texas at Austin, Austin, TX 78712, USA. xc4646@utexas.edu
4. Allen Institute for AI (AI2), Seattle, USA davida@allenai.org
5. IBM Research Yorktown Heights, USA, adhuran@us.ibm.com

* Corresponding author: saleh.afroogh@utexas.edu

## Abstract

The integration of Generative Artificial Intelligence (GAI) and Large Language Models (LLMs) in academia has spurred a global discourse on their potential pedagogical benefits and ethical considerations. Positive reactions highlight some potential, such as collaborative creativity, increased access to education, and empowerment of trainers and trainees. However, negative reactions raise concerns about ethical complexities, balancing innovation and academic integrity, unequal access, and misinformation risks.

Through a systematic survey and text-mining-based analysis of global and national directives, insights from independent research, and eighty university-level guidelines, this study provides a nuanced understanding of the opportunities and challenges posed by GAI and LLMs in education. It emphasizes the importance of balanced approaches that harness the benefits of these technologies while addressing ethical considerations and ensuring equitable access and educational outcomes. The paper concludes with recommendations for fostering responsible innovation and ethical practices to guide the integration of GAI and LLMs in academia.

# I. Introduction and Background

Recent developments and deployments of artificial intelligence (AI) in academia, specifically generative AI (GAI) like Large Language Models (LLMs), have raised both excitements and concerns. GAI draws on patterns learned from large datasets to engender new content, including texts, images, and music. LLMs, as prominent examples of GAI, are sophisticated neural network architectures designed to understand and predict probability distributions within linguistic sequences, fundamentally altering how we interact with and produce written content.

The field of Natural Language Processing (NLP) has witnessed significant advancements with the emergence of GAI and LLM. LLMs are essential tools within NLP, designed to understand and predict probability distributions within sequences of linguistic units such as words or phrases. These models, characterized by their vast number of parameters ranging from tens of millions to billions, undergo extensive training using large volumes of diverse data [1]. This training process enables LLMs to grasp intricate linguistic patterns, improving their ability to generate coherent and contextually appropriate text. Functioning in an autoregressive manner, LLMs predict the probability distribution of a token based on its preceding tokens, leveraging the chain rule of probability and conditional probabilities [2]. The generative nature of LLMs allows them to produce human-like text, embodying sophisticated neural architectures that contribute to their contextual understanding, semantic coherence, and adaptability. These advancements in LLMs and GAI have profound implications across various domains, including academia, where their potential impact on teaching, research, and knowledge dissemination is increasingly recognized [1], [3].

1.1: Escalating significance and prevalence of LLMs and GAIs in academic settings

Within academia, the integration of LLMs and other GAI technologies has introduced unprecedented opportunities and challenges. These models have shown remarkable capabilities in generating human-like text, aiding in tasks such as automated essay grading, content creation, and language translation. However, their usage also raises significant pedagogical, ethical, and legal considerations that necessitate careful evaluation and guidelines for responsible deployment.

The increasing relevance and widespread use of GAI/LLM in academic contexts have prompted diverse responses and adaptations from universities worldwide. As evidenced by the actions taken by Australian universities, banning AI tools outright may not be a feasible solution due to students' ability to circumvent such restrictions using technologies like VPNs. Consequently, some institutions have revisited their assessment methods, opting for traditional pen-and-paper exams to maintain academic integrity [4].

Moreover, the global landscape reflects a spectrum of responses regarding the use of GAI in academic settings. Many universities have categorically labeled the use of AI tools for assignments and exams as academic misconduct, while others have revised their plagiarism policies and AI-use guidelines to address these challenges. In a BestColleges survey, approximately 31% of participants reported restrictions imposed by instructors, course materials, or school honor codes regarding the use of AI tools [4]

Despite the growing adoption of GAI applications in education, a UNESCO global survey highlights a critical gap: fewer than 10% of schools and universities have developed formal policies or guidance specific to the use of such technologies[5]This lack of institutional preparedness underscores the urgent need for comprehensive frameworks that balance innovation with ethical considerations in leveraging LLMs and other GAI tools within academia.

This paper seeks to delve into the multifaceted landscape of GAI, with a primary focus on LLMs, within academic settings. Through a careful consideration of the pedagogical implications, ethical considerations, and legal frameworks surrounding the use of these powerful AI tools, this study aims to provide insights and recommendations for navigating the complex intersection of AI technology and academia. Through a systematic evaluation of existing guidelines and practices, we aim to contribute to the development of comprehensive and ethically sound frameworks for the usage of GAIs, including LLMs, in academic contexts.

## II. Methodology

In this study, we conducted a comprehensive comparative analysis based on data collected from 80 universities to evaluate the pedagogical, ethical, and legal guidelines for using GAIs and LLMs in academic settings. Our approach was methodologically rigorous, ensuring a thorough examination of the guidelines. We carefully selected universities to represent a diverse range of institutions globally, including top-ranked universities from six continents recognized for their academic excellence. Our inclusion criteria also considered different types of universities, such as humanities, technological, public, and private institutions, to capture a broad spectrum of perspectives.

In the evaluation of academic guidelines for GAI/LLM, selecting appropriate categorization criteria is paramount for gaining a nuanced understanding of the diverse recommendations landscape. While there are several potential criteria such as Academic Level, Type of Institution, and Disciplinary Focus, two key criteria that emerge for comprehensive coverage are geographical location and operational levels. Geographical location is crucial as it illuminates the influence of regional factors, legal frameworks, and cultural contexts on guideline development and application. This dual approach enables a holistic exploration, encompassing both specialized insights from academic disciplines and broader global perspectives that shape GAI/LLM guidelines.

We emphasized the importance of diversity within each continent, recognizing the influence of cultural backgrounds and regional factors on the development and implementation of guidelines related to GAI and LLM. Additionally, we referred to relevant departmental guidelines from various regions, such as European and Asian countries, regarding the use of specific AI tools like ChatGPT, ensuring a comprehensive and inclusive analysis. This meticulous selection process and broad data collection approach enabled us to conduct a robust comparative analysis and derive meaningful insights into the varying approaches and considerations guiding the use of GAI and LLM in academia globally.

Furthermore, operational levels, specifically at the university and departmental levels, play a pivotal role in understanding how guidelines are implemented and adapted within organizational structures. This understanding allows for an examination of practical aspects, challenges, and variations in guideline implementation, ensuring a comprehensive assessment of their impact on different operational levels within academic institutions. By emphasizing geographical location and operational levels as primary categorization criteria, our study aims to provide a comprehensive and nuanced analysis of academic guidelines for GAI and LLM, taking into account regional dynamics and organizational contexts.

In the data collection phase, we initially focused on identifying universities with official guidelines specifically addressing the usage of GAI, with a primary emphasis on LLMs. This targeted approach aimed to maintain methodological consistency and reliability. Universities lacking official guidelines were deliberately excluded from our study to ensure the credibility of our analysis. To compensate for these exclusions, replacement universities were carefully selected based on the same criteria of top-tier status, geographic diversity, and cultural representation. This meticulous selection process guaranteed a balanced and representative sample, enhancing the validity of our findings (See, Table 1.)

Following data collection, our literature review process involved a systematic categorization and comprehensive analysis of the collected guidelines from the selected universities. Each guideline underwent rigorous scrutiny concerning its scope, recommendations, ethical considerations, and pedagogical implications related to the usage of LLMs and GAI in academic contexts. This meticulous examination enabled us to identify major themes, common trends, and areas of divergence among the guidelines, providing a holistic understanding of the landscape of guidelines in this domain.

Subsequently, we conducted the text mining-based analysis of the guidelines. We initiated the analysis with the tokenization of the text data collected from 80 guidelines. Tokenization involved breaking down the bulk of text into sentences and then into individual words, which enabled us to scrutinize the text at a granular level. We employed the Natural Language Toolkit's (nltk) sent_tokenize and word_tokenize functions for this purpose. To refine our data further and to eliminate noise and focusing on substantive text, we filtered out stopwords—commonly used words that add little semantic value—using nltk's stopwords list. This step was used to eliminate noise and focusing on substantive text.

Following tokenization, we implemented stemming and lemmatization. Stemming, via nltk's PorterStemmer, reduced words to their root form, thereby grouping different forms of the same word. Lemmatization, a more context-aware process performed by WordNetLemmatizer, converted words into their dictionary form. Both methods were instrumental in normalizing textual data, ensuring that variations of a word are treated as a single item. This approach enhances the accuracy of frequency counts in the analysis.

Afterwards, with a cleansed and normalized dataset, we constructed a Term Frequency-Inverse Document Frequency (TF-IDF) model using sklearn's TfidfVectorizer. This model assessed the importance of a word relative to a document within a corpus of documents, by balancing the word frequency (Term Frequency) against the rate at which the word appears across multiple documents (Inverse Document Frequency). The TF-IDF model was used in identifying words that are unique and significant to specific documents, which, in our case, helped highlight key terms within the academic guidelines text. Further, we utilized the KMeans algorithm for clustering, which grouped words into clusters based on their TF-IDF scores, revealing patterns and themes within the text. Complementing this, used for clustering visualizations illustrated the prominence of different terms.

**Table 1.** Workflow of the analysis of GAI/LLM guidelines

| Analytics Workflow | Description | |
| --- | --- | --- |
| Data preparation | Focus on top-tier institutions with diverse representation globally. | Identification of universities with official guidelines on GAI and LLMs. |
| Data preprocessing | Exclusion of universities without official guidelines. | Selection of replacement universities meeting inclusion criteria. |
| Systematic review | Systematic categorization and comprehensive analysis of collected 80 guidelines. | Scrutiny of scope, recommendations, ethical considerations, and pedagogical implications. |
| Qualitative analysis | Identification of major themes, minor codes, common trends, and areas of divergence among guidelines. | Emphasizing mitigation strategies and proposing roadmap |
| Quantitative Exploration | Analyzing text through tokenization using NLTK's sent_tokenize and word_tokenize functions. Filtering out stopwords to focus on substantive text. | |
| | Implementing stemming with NLTK's PorterStemmer and lemmatization with WordNetLemmatizer to normalize textual data. | |
| Text mining-based Analysis | Constructing a TF-IDF model using sklearn's TfidfVectorizer to assess word importance relative to documents. | |
| | Using KMeans clustering to group words based on TF-IDF scores, revealing patterns and themes. | Utilizing clustering visualizations to highlight differences |

# III. Qualitative Findings and Resultant Themes

## 3.1. Global and cross-national directives for usage of LLMs and GAIs

In the realm of academia, the integration of GAI/LLM has sparked a global discourse centered on their potential benefits and concerns. This discourse is reflected in a spectrum of directives and reactions at both the global and national levels, shaping the acceptance and reception of these advanced technologies within educational frameworks. While some directives emphasize the positive impacts and opportunities afforded by GAI/LLM, others underscore cautionary considerations regarding fairness, privacy, and equitable access. These divergent perspectives are echoed in independent research findings, highlighting a nuanced landscape where optimism about technological advancements coexists with apprehensions about their societal and educational implications.

### 3.1.1. Global and National Acceptance and Reception (Positive Reaction) to Usage of LLMs and GAIs in Academia:

In general, the trend in academia seems to be toward acknowledging potential issues while providing guidance to maximize the benefits of GAI in education while minimizing the harms. For example, some systems [6], [7] focus almost entirely on how to make GAI advantageous without explicitly acknowledging potential drawbacks. This includes guidelines on how to write effective prompts, delete chat history, and to ensure that educators test the GAI systems prior to using them in class. Many other directives strike a more balanced approach, but expressly note positive uses of GAI. Some notable uses include the potential for GAI to be more adaptive to students [8], [9], sometimes expressed as personalized learning [[8], [9], [10], [11]. Going further, some nations see GAI as a potential tutor, not only adapting lessons based on student understanding, but also helping to lead the students through the lessons[8], [9], [11]. Other potential uses include assisting educators with administrative tasks[8], [11], increasing access to education to students beyond what a human teacher can reasonably provide [8], assisting with professional development [8], and helping to provide individualized feedback to students and educators [8], [9], [11].

3.1.2. Denial, Banning, Concerns (Negative Reactions) of Usage of LLMs and GAIs in Academia:

It does not seem that any country has a national directive banning the use of GAI in academia, aside from countries that lack access to, or ban the use of, GAI in general, such as North Korea. However, several directives urge educators to consider potential harms of GAI in education. One prominent set of concerns centers around bias and fairness, where directives note the potential of GAI to favor certain viewpoints over others[8], [9], [10], [12] Just as often, directives note the potential for privacy violations based on content students and educators may include in prompts [8], [9], [10], [11], [12]. The third most frequently cited concern raised the issue of potential unequal or inequitable access to GAI due to limited access to computers, the internet, and/or knowledge of how to utilize these novel tools for learning [8], [9], [10], [11], [12], [13].

Some less frequently cited concerns included the potential for reliance on GAI to lead to less social interaction and collaboration [8], [12], and the potential for overdependence on the technology rather than learning and internalizing new knowledge and skills [8], [9], [12]. Finally, there are some educator-specific concerns, such as the potential of educators to lose their autonomy in the classroom as they and their students turn to GAI to guide lessons [8], [9], [10], [12], educators being unprepared and untrained to make meaningful use of GAI [8], [12], and a general fear that GAI will undermine the respect and value educators receive[8].

3.1.3. Perceptions of LLMs and GAIs in Education: More Insights, Independent Research

Perhaps unsurprisingly, the amount of high-quality peer-reviewed research around LLMs and education is limited given the novelty of the technology. However, we will note two studies here. First, the findings above, derived from reviewing national directives, align with the mixed sentiments discovered in independent research that examined sentiment regarding ChatGPT and its use in education on Twitter. The study found that about 40% of the relevant tweets were about the "opportunities, limitations, and consequences of ChatGPT [in education]" and "efficiency and cheating when students use ChatGPT to write." The authors found that sentiment overall sorted into a mix of 40% positive, 30% negative, and 30% neutral. The topic with the largest negative sentiment was the capability of ChatGPT to replicate student papers, which could lead to cheating and other undesirable outcomes. The most positive topic was about how ChatGPT could lower the cost of education, presumably by making knowledge more accessible and scalable.[14] A separate study involved interviewing college students, college faculty, and education experts to assess their views on ChatGPT. It found a similar mixed sentiment about the technology, noting that the technology is both quick and easy to use among other possible benefits, while acknowledging that it could lead to students and faculty over-relying on it and it could undermine academic integrity. [15]

### 3.2. University-level Guidelines for usage of LLMs and GAIs

In navigating the dynamic landscape of integrating GAI/LLM into academic settings, universities play a pivotal role in formulating and implementing guidelines that govern their usage. Recognizing the transformative potential of these technologies in enhancing teaching, research, and knowledge dissemination, universities worldwide are faced with the dual imperative of harnessing their benefits while mitigating potential risks. This section explores the diverse array of university-level guidelines tailored to regulate the utilization of GAI/LLM within educational contexts. Drawing from an extensive analysis of 80 official guidelines sourced from universities across different countries in six continents with diverse cultures and approaches, this study provides a comprehensive examination of the global landscape of AI governance in academia (See Table 2).

**Table 2: University-level guidelines for Usage of LLMs and GAIs**

|   | Continent | Country | University | Name of document/website |
|---|---|---|---|---|
| 1 | Africa | South Africa | University of Cape Town | Artificial Intelligence for Teaching & Learning [16] |
| 2 | | Egypt | Cairo University | FCAI Policy and Guidelines for use of Generative AI in Postgraduate Studies and Research[17] |
| 3 | | South Africa | University of Witwatersrand | ChatGPT & other AI tools for Learning and Teaching[18] |
| 4 | | South Africa | University of Pretoria | Leveraging Generative Artificial Intelligence for Teaching and Learning Enhancement[19] |
| 5 | | South Africa | University of Johannesburg | UJ Practice Notes: Generative Artificial Intelligence in Teaching, Learning and Research[20] |
| 6 | | South Africa | Stellenbosch University | Stellenbosch University Academic Integrity: Responsible Use of AI tools[21] |
| 7 | | South Africa | University of the Free State | Stepping up with ChatGPT - AI-assisted Technology in Education[22] |
| 8 | | Egypt | The American University in Cairo | AUC's Statement on the Use of Artificial Intelligence Tools[23] |
| 9 | | South Africa | North-West University | Implications of AI for teaching and learning in higher education & Guidelines for the Utilization of AI in Teaching and Learning at NWU[24] |
| 10 | | South Africa | African Observatory on Responsible Artificial Intelligence | Generative AI guidelines at South African universities[25] |
| 11 | North America | USA | Harvard | Guidelines for Using ChatGPT and other Generative AI tools at Harvard[26] |
| 12 | | USA | Stanford | Generative AI Policy Guidance[27] |
| 13 | | USA | MIT | Getting Started with AI-Enhanced Teaching: A Practical Guide for Instructors[28] |
| 14 | | USA | Princeton | Generative AI Guidance[29] |
| 15 | | USA | University of Chicago | Guidance for Syllabus Staatements on the Use of AI Tools[30] |
| 16 | | USA | Columbia | Considerations for AI Tools in the Classroom[31] |
| 17 | | USA | California Institute of Technology | Guidance on the Use of Generative AI and Large Language Model Tools[32] |
| 18 | | USA | University of California, Berkeley | Appropriate use of ChatGPT and Similar AI Tools[33] |
| 19 | | USA | Yale | Guidelines for the Use of Generative AI Tools[34] |
| 20 | | USA | University of Pennsylvania | Statement on Guidance for the University of Pennsylvania Community on Use of Generative Artificial Intelligence[35] |
| 21 | | USA | University of California, Los Angeles | Guidance for the use of generative AI[36] |
| 22 | | USA | Cornell University | Cornell Gudelines for artificial intelligence[37] |
| 23 | | Canada | University of Toronto | Generative Artificial Intelligence in the classroom[38] |
| 24 | | Canada | University of British Columbia | Generative AI – Academic Integrity at UBC[39] |
| 25 | | Canada | McGill University | Principles on Generative AI in Teaching and Learning at McGill[40] |
| 26 | | Canada | University of Alberta | AI-Squared – Artificial Intelligence and Academic Integrity[41] |
| 27 | | Canada | University of Waterloo | Artificial Intelligence and ChatGPT – Academic Integrity[42] |
| 28 | | Canada | University of Montreal | Montreal Declaration on Responsible AI[43] |
| 29 | | Canada | McMaster University | Provisional Guidelines on the Use of Generative AI in Teaching and Learning[44] |
| 30 | | USA | U.S. Department of Education | Artificial Intelligence and the Future of Teaching and Learning[45] |
| 31 | South America | Colombia | Universidad del Rosario | Guidelines for the Use of Artificial Intelligence in University Courses[46] |
| 32 | | Argentina | University of Buenos Aires | Guidelines for the use of ChatGPT and text generative AI in Justice[47] |
| 33 | | Colombia | Universidad de Los Andes | Guidelines for the use of artificial intelligence in university contexts[48] |
| 34 | | Colombia | Pontifical Javeriana University | Editorial Policy, Publication Ethics and Malpractice Statement[49] |
| 35 | | Argentina | Universidad de san andres | Readiness of the judicial sector for artificial intelligence in Latin America[50] |
| 36 | | Peru | Government and Digital Transformation Secretariat | National Artificial Intelligence Strategy[51] |
| 37 | | Chile | Ministry of Science, Technology, Knowledge and Innovation | Guidelines for the use of artificial intelligence tools in the public sector[52] |
| 38 | | Chile | Pontificia universidad católica de chile | ChatGPT: How to use it in classes?[53] |
| 39 | Asia | China | Tsinghua University | International AI Cooperation and Governance Forum 2022[54] |
| 40 | | Singapore | National University of Singapore | Responsible Use of AI – Guidance from a Singapore Regulatory Perspective[55] |
| 41 | | Japan | Nagoya University | Regarding the Use of Generative AI[56] |
| 42 | | Japan | University of Tokyo | Guidelines for Instructors Regarding AI in University Education at Tokyo University of Foreign Studies[57] |
| 43 | | Hong Kong SAR | University of Hong Kong | Use of Artificial Intelligence Tools in Teaching, Learning and Assessments: A Guide for Students[58] |
| 44 | | South Korea | Seoul National University (SNU) | Seoul National University AI Policy Initative[59] |
| 45 | | Japan | Kyushu University | Note on the Use of Generative AI in Education at Kyushu University – For Teachers -[60] |
| 46 | | Japan | Weseda University | About the Use of Generative Arificial Intelligence (ChatGPT, etc.)[61] |
| 47 | | South Korea | Ulsan National Institute of Science and Technology | A Guide to the Use of Generative AI[62] |
| 48 | | Singapore | Signapore Management University | SUM Framework for the use of Generative AI Tools[63] |
| 49 | | Singapore | Singapore University of Technology and Design | Artificial Intelligence in Education[64] |
| 50 | | Singapore | Singapore Institute of Technology | Generative AI at the Singapore Institute of Technology[65] |
| 51 | | Singapore | Nayang Technological University | NUT Position on the Use of Generative Artificial Intelligence in Research[66] |
| 52 | | Taiwan | National Tsing Hua University | Guidelines for Collaboration, Co-learning, and Cultivation of Artificial Intelligence Competencies in University Education[67] |
| 53 | | Taiwan | National Taiwan University | Guidance for Use of Generative AI Tools for Teaching and Learning[67] |
| 54 | | Hong Kong SAR | The Chinese University of Hong Kong | Use of Artificial Intelligence Tools in Teaching, Learning and Assessments A Guide for Students[68] |
| 55 | | Thailand | Chulalongkorn University | Chulalongkorn University Principles and Guidelines for using AI Tools[69] |
| 56 | | Malaysia | Universiti Malaya | ChatGPT General Usage[70] |
| 57 | | Malaysia | Universiti Putra Malaysia | Guide for ChatGPT usage in Teaching and Learning[71] |
| 58 | | China | The Supervision Department of the Ministry of Science and Technology | Guidelines for Responsible Research Conduct (2023)[72] |
| 59 | Australia | Australia | The Department of Education | The Australian Framework for Generative Artificial Intelligence (AI) in Schools[73] |
| 60 | | New Zealand | The University of Auckland | Advice for students on using generative artificial intelligence in coursework[74] |
| 61 | | UK | University of Oxford | Use of generative AI tools to support learning[75] |
| 62 | | UK | University of Cambridge | Artificial intelligence and teaching, learning, and assessment[76] |
| 63 | | UK | Imperial College London | Generative AI Guidance[77] |
| 64 | | UK | London School of Economics and Political Science | School Statement on Generative Artificial Intelligence and Education[78] |
| 65 | | UK | University College London | Using AI tools in assessment [79] |
| 66 | | UK | The University of Edinburgh | AI Guidance for Staff and Students[80] |
| 67 | Europe | Netherlands | Erasmus University Rotterdam | Gnenerative AI in education[81] |
| 68 | | Belgium | KU Leuven | Responsible use of generative Artificial Intellgence[82] |
| 69 | | Switzerland | ETH Zurich | AI in education, resources for teaching faculty[83] |
| 70 | | Netherlands | University of Amsterdam | AI tools and your studies[84] |
| 71 | | Norway | University of Oslo | Guidelines to use artificial intelligence at UiO[85] |
| 72 | | Finland | University of Helsinki | Using AI to support learning[86] |
| 73 | | Italy | University of Padua | Research in Artificial intelligence[87] |
| 74 | | Sweden | Stockholm University | Guidelines on using AI-powered chatbots in education and research[88] |

| 75 | | Denmark | Technical University of Denmark | [DTU opens up for the use of artificial intelligence in teaching](#)[89] |
| 76 | | Netherlands | Delft University of Technology | [AI chatbots in unsupervised assessment](#)[90] |
| 77 | | Portugal | Universidade de Lisboa | [Artificial Intelligence in education – Técnico presents resolution on the use of tools such as ChatGPT](#)[91] |
| 78 | | Netherlands | University of Utrecht | [Guidelines for the use of generative AI](#)[92] |
| 79 | | Switzerland | University of Zurich | [Guidelines for the Use of AI Tools](#)[93] |
| 80 | | UK | Russell Group (24 UK research-intensive universities) | [Russell Group principles on the use of generative AI tools in education](#)[94] |

The results of this exploration are categorized into the following nine subsections, each addressing key aspects of AI regulation and ethical practice within higher education institutions (see Table 3). By delving into the strategies employed by universities to navigate the opportunities and challenges posed by GAI/LLM, this section aims to shed light on the multifaceted nature of AI governance within academia and offer insights for fostering responsible innovation and ethical practice.

### 3.2.1. Prioritizing Responsibility and Safety: Two Imperatives in GAI/LLM Integration

Responsibility and safety are two key requirements of an ethical use of AI systems in integrating GAI and LLMs into academic settings. From the perspective of academic institutions, critical or original thinking is a significant academic value that should not be compromised by irresponsible use of AI systems or LLMs, which may even lead to cases of plagiarism [17], [18], [20], [34], [35], [37]. For this reason, they expect students to be transparent about their use of AI tools [20], [31], [34], while safeguarding confidential data [19], [38], [62], [66]. Furthermore, considerations of responsibility require students to explore alternative tools that prioritize privacy [19], [20], [37], [39].

Responsibility in GAI usage entails acknowledging the tools used and addressing factual errors or fabricated references they may generate [18], [20], [26]. Academics bear the responsibility of training students in appropriate AI use, highlighting AI limitations, and emphasizing human intervention in research [18], [20], [31], [38]. Instructors play a pivotal role in teaching about AI and managing related course policies [39],[44],[45]. Students must understand citation expectations when using AI [64], [67].

Safety concerns focus on ensuring that AI systems are safe, interpretable, and robust [45], [72], [73] adapted to educational contexts to enhance trust [45], [72], [85]. Data used to train AI systems should not involve personal or sensitive information to prevent leaks [26], [33], [62]. Cautionary directives also advise against using AI tools for sensitive data due to potential privacy risks [16], [20], [47]. These measures collectively promote responsible, safe, and privacy-conscious AI integration in academia.

### 3.2.2. Navigating Ethical Complexities and Human-centric Usage of GAI/LLM

Addressing the ethical dilemmas posed by AI and LLMs in academia is intricate. Institutions are addressing concerns such as fairness, privacy, and accessibility, recognizing both the promise and the peril of these technologies [28], [36], [44], [92], [94]. There's concern that AI/LLMs might perpetuate biases and spread misinformation due to biased training data [16], [18], [29], [88], [92]. To address this, institutions stress the need to critically assess AI-generated content, diversify training data, and maintain human oversight through a "human-in-the-loop" approach [45]. Additionally, ensuring equal access to technology and AI tools is crucial for fairness among students [16], [45], [59], [72].

On the ethical front, concerns revolve around biases in AI, including cognitive and gender/racial biases, as well as privacy issues related to data gathering and leakage [47], [62] [20]. Accessibility and equity are also highlighted, noting the impact of the digital divide on students' access to advanced AI tools [16], [19], [36]. A human-centric approach is advocated to ensure AI supports learning without overshadowing students' cognitive efforts [45]. Institutions are urged to develop clear policies on data protection and ethical AI usage while promoting research into open-source and affordable AI alternatives [28], [49], [55].

### 3.2.3. Balancing Innovation and Integrity: Strategies for Ethical AI/LLM Integration

Balancing innovation with integrity in the integration of AI and LLMs within academic settings requires a nuanced approach that addresses both the potential benefits and ethical concerns. Institutions navigate this landscape by implementing strategies that balance the use of these technologies while upholding academic integrity standards. A common strategy involves prohibiting unauthorized AI/LLM usage for generating academic work, emphasizing the importance of originality and critical thinking [17], [18], [20], [63], [75], [80]. However, recognizing the value of AI/LLMs in certain contexts, institutions adopt measures such as restricted usage, specific stages for AI integration (like brainstorming or editing), and promoting critical evaluation of AI-generated content by students [16], [20], [86]. Moreover, alternative assessment methods like oral exams and project-based tasks are explored to mitigate potential misuse of AI [18], [45], [60], [62], [88]. In the research domain, institutions may impose restrictions on AI usage for topics or methodologies with ethical concerns or risks [67], [72].

Despite the advantages, integrating AI and LLMs faces challenges such as limited internet access affecting real-time data retrieval, input text constraints, potential biases, lack of context, and creative limitations [16]. While not universally banned, concerns over academic integrity prompt universities to carefully consider AI tool usage within courses, emphasizing clear expectations and responsible AI utilization [16], [45], [94]. These deliberations reflect a proactive stance in harnessing AI's potential while safeguarding academic integrity, highlighting the ongoing need for ethical guidelines and thoughtful implementation of AI/LLMs in educational environments.

**Table 3**: Major and minor codes included in academic general guidelines

|   | **Major codes (Categorical Themes)** | **Minor codes (Keywords/Key Concepts)** |
|---|---|---|
| 1 | **Responsibility & Safety** | Nurturing critical thinking, Transparency, Confidential information protection, Privacy features, Security features, Independent critical thinking, AI tool usage disclosure, Unauthorized use condemnation, Plagiarism prevention, Cheating prevention |
| 2 | **Ethical complexities & Human-centric usage,** | Ethical complexities, Human-centric usage, Fairness concerns, Privacy concerns, Accessibility concerns, Societal biases, Discriminatory outputs, Misinformation, Critical evaluation, Diversifying training data, Human oversight, Human-in-the-loop approach, Unequal access, Digital divide |
| 3 | **Balancing Innovation and Integrity** | Ethical AI/LLM integration, Limitations, Restrictions, Responsible integration, Academic integrity, Proper disclosure, Critical evaluation, Reflective usage, Alternative assessment methods, Oral exams, Project-based tasks, Research restrictions, Risk management. |
| 4 | **Truth assurance & Misinformation risks** | GAI/LLM outputs, Reviewing and verifying content, Accuracy and reliability, Knowledge formation, Plausible-sounding outputs, Factual accuracy, Fact-checking, Skepticism fostering, Transparent AI models, Accountable AI models, Reasoning transparency, Bias detection, Bias mitigation, Diversifying training data. |
| 5 | **Pedagogical innovation & GAI/LLM literacy** | Pedagogical innovation, GAI/LLM literacy, Educational settings, Learning experiences, Personalized instruction, Knowledge creation, Exploration tools, Critical thinking skills, Brainstorming, Creative content generation, Feedback provision, Engaging learning experiences, Independent learning skills, Discerning accurate information, Human oversight. |
| 6 | **Collaborative creativity & Co-designing** | Co-authoring processes, Content creation, Editor role, Research assistant role, Content refinement, Teaching assistant role, Self-study assistant role, Personalized learning paths, Coding assistance, Summarizing, Content development, Skill democratization, Scalable tutoring systems, Intelligent tutoring systems, Continuous learning |
| 7 | **Empowering educators/staff/faculty** | Evaluation methodology, Grading practices, Fair/Efficient assessment, Educational management, Administrative processes, Predictive analytics, Syllabi creation, Research proposal creation, Teaching tool, Intelligent tutoring systems, Virtual assistants, AI-enabled educational technologies, Equitable education, Efficient education |
| 8 | **Empowering students' study/research** | Responsible use, Academic honesty, Research methodologies, Critical thinking, Human oversight, AI-assisted research, Brainstorming, Drafting, Revising work, Deep engagement, Content understanding, Personal growth, Feedback loops, Self-reflection, Curiosity, AI-integrated education, Digital learning landscape |
| 9 | **Tailored Guidance & Specific Guidelines** | Tailored guidance, Safety Guidelines, Educational Guidelines, Ethical Guidelines, Discipline-Specific Guidelines, Student Guidelines, Educator and Faculty Guidelines, Staff Guidelines, Digital learning environment, Personalized learning, Teaching methodologies, Industry-specific skills, Harmonious balance, Stakeholder guidance. |

### 3.2.4. Ensuring Truth in GAI/LLM Outputs and Addressing Misinformation Risks

Ensuring Truth and avoiding misinformation in GAI/LLM Outputs requires a multidimensional approach due to the persistent concern of AI and LLMs generating inaccurate or fabricated information, also known as *hallucinations*. The issues arise from how generative AI models tend to generate outputs that sound more plausible over being factually correct. This tendency has resulted in the proliferation of misinformation and deceptive assertions [16], [47], [62]. To address this, institutions recommend critical review, evaluation and verification of the generated content as well as fact-checking against reliable sources, and fostering skepticism among users [50], [90]. Transparent and accountable AI models are also explored to enable users to understand the reasoning behind outputs and assess their validity [49], [65], [90]. Additionally, strategies such as diversifying training data and implementing bias detection and mitigation are crucial in reducing the potential for biased AI-generated content [16], [47], [62].

Specific challenges like *hallucination* and the risk of spreading misinformation are central to the discussion. Hallucination occurs when AI systems provide answers that are statistically likely but not factually accurate, highlighting the need for post-processing, model fine-tuning, and improved data quality to mitigate this issue. Misinformation, closely linked to hallucination, underscores the importance of user responsibility in verifying accuracy, monitoring reliability, and addressing educational implications related to AI outputs [16], [47], [62], [62]. Despite the capabilities of Generative AI and LLMs, addressing these challenges requires advancements in technology, user education, and ethical considerations to ensure responsible and accurate use of these powerful tools.

### 3.2.5. Empowering Pedagogical Innovation through GAI/LLM Literacy

Enabling Innovative Teaching Methods with AI/LLM Literacy involves a comprehensive strategy that recognizes the significant impact of AI and LLMs in education. This approach emphasizes responsible usage and the development of critical thinking skills. Some institutions' guidelines recognize the potential of generative AI technologies to enhance unique learning opportunities, personalize teaching, and facilitate knowledge formation [29], [45], [51], [68]. However, cultivating AI literacy among students and educators plays a central role in empowering them to ensure a nuanced understanding of AI/LLMs' capabilities, biases, and the necessity for critically evaluating AI-generated content [16], [18], [29], [45], [57], [77].

Professional development programs are advocated to integrate AI literacy deeply into teacher professionalism, enabling educators to harness AI tools innovatively and foster trust through transparency [16], [18], [29], [45], [57], [68]. Moreover, the integration of AI in educational tasks aims to enhance creativity and critical thinking, with strategies emphasizing prompt engineering, reflection on AI-generated outputs, and reevaluation of traditional notions of originality and creativity within academic contexts [18], [28], [36], [37], [49], [62] , [64] This balanced approach not only acknowledges AI's potential to transform educational practices but also underscores the importance of ethical usage, critical thinking, and human oversight in shaping a thoughtful pedagogical strategy for the AI era.

### 3.2.6. Collaborative Creativity: Co-Designing with AI for Enhanced Content and Creative Workflows

The integration of AI in co-designing and co-authoring processes [16], [18], [20], [45], [46], [48], [49] has shown to enhance creativity and efficiency in content creation, its role as an editor and research assistant [18], [32], [41], [49], [62], [72] offers precision and depth in content refinement and research methodologies. As a teaching and self-study assistant, AI's adaptability, and personalized learning paths[36], [41], [44], [45], [91]significantly contribute to individual learning experiences and outcomes. Its application in coding [19], [62], [63], [75], [80], [90], summarizing, and developing [18], [29], [31], [47], [48], [61] , [67], [67] extends its utility to technical and non-technical domains alike, offering tools that democratize these skills. Leveraging AI, providing scalable intelligent tutoring systems, and personalized learning experiences [16], [19], [20], [45], [67], are pivotal in educational settings. Guidelines advocate for a balanced approach between technological advancements and human-centric values in education, ensuring equitable access and fostering a culture of continuous learning and improvement.

3.2.7. Empowering Educators: Guidelines for Staff and Faculty in Leveraging LLMS and GAIs

The guidelines suggest a refined methodology for evaluation, grading, and assignments [44], [45], [47], [48], [53], [57], [69], highlighting the potential of AI for fair and efficient assessment practices. They advocate for enhanced communication channels between staff and students, as well as amongst staff members [16], [27], [30], [85]. Within the realm of educational management, the guidelines propose integrating AI systems for streamlined administrative processes, predictive analytics for student performance, and personalized learning experiences [24], [45], [53], [72]. The creation of syllabi [37], [38], [67]and research proposals [17], [66] is offering staff and faculty novel tools to design more engaging and relevant educational materials. Furthermore, the guidelines underscore the utilization of AI as a teaching tool, encouraging the adoption of intelligent tutoring systems, virtual assistants, and other AI-enabled educational technologies[16] , [23], [28], [45], [62], [67], [83], [85]. Collectively, these guidelines aim to empower educators with AI tools and insights, fostering an educational landscape that is equitable, efficient, and aligned with the evolving needs of learners and educator.

3.2.8. Empowering students: Guidelines for Research and Assignments with LLMS and GAIs

The guidelines encourage students to harness AI technologies responsibly, ensuring that their use in academic work adheres to principles of honesty and originality, while also leveraging these tools for enhancing research methodologies and insights [16][17], [18], [19], [21], [36], [39], [42], [44]. They advocate for a balanced approach to interpreting and applying AI-generated data and conclusions, emphasizing the importance of critical thinking and human oversight in AI-assisted research [16], [20], [23], [24], [25], [35], [37], [45], [72]. In assignments, the guidelines suggest leveraging AI for brainstorming, drafting, and revising work, yet they stress the necessity of students' deep engagement with content to cultivate understanding and personal growth [16], [18], [29], [36], [62]. The guidelines also highlight the potential of LLMS and GAIs to provide students with feedback loops for self-reflection, motivation, discovery and assessment [18], [19], [20], [24], [31], [45], [67], [72]. By adhering to these guidelines, students are encouraged to navigate the digital learning landscape with integrity, curiosity, and a commitment to personal and academic excellence, positioning themselves as adept learners in an increasingly AI-integrated educational environment.

### 3.2.9. Tailored Guidance: Navigating AI Integration with Specific Guidelines

Safety guidelines are paramount, focusing on protecting users from harm and securing data against breaches, thus ensuring a safe digital learning environment [24], [45], [50], [73], [85], [92]. Educational guidelines aim to enhance teaching and learning experiences, advocating for pedagogical practices that integrate AI to support personalized learning, engagement, and academic success [16], [20], [24], [45], [49], [61], [53], [70]. Ethical guidelines underscore the importance of transparency, fairness, and accountability in AI applications, emphasizing the prevention of bias and the promotion of equitable access to educational technologies [20], [35], [45], [47], [48], [49], [51], [55], [67]. Department-specific guidelines recognize the unique requirements of different fields of study, offering tailored advice on leveraging AI for discipline-related research, teaching methodologies, and industry-specific skills development [18], [35], [72], [77], [93]. These guidelines collectively aim to foster environments where safety, educational quality, ethical standards, and disciplinary relevance are harmoniously balanced, guiding stakeholders in navigating the complexities of integrating AI into the academic fabric.

# IV. Quantitative Findings and Resultant Patterns

We conducted a text mining-based analysis consisting of models and techniques—from tokenization to visualization. Using these models, we achieved a detailed analysis of the academic guidelines. Each step, from the initial tokenization of text to the final visualization of data, was to distill the information contained within the guidelines. This process provided result informed by raw frequency and the contextual significance of each term, a foundation for the development of more comprehensive academic guidelines.

### 4.1. Quantitative Frequency Analysis

The initial phase involved a quantitative frequency analysis. We scrutinized the data using a series of nine plots, each representing different categorical themes (i.e., major code in table 3) related to academic guidelines. These categorical themes ranged from "Responsibility & Safety" to "Tailored Guidance & Specific Guidelines." The analysis quantified the prevalence of specific keywords within the text corpus, translating raw data into visual representations. Figure 1 depicted the frequency of keywords such as "privacy," "plagiarism," etc., within nine categorical themes, offering a clear depiction of the most and least mentioned keywords.

**Figure 1**: The Frequency of Keywords in the Nine Major codes

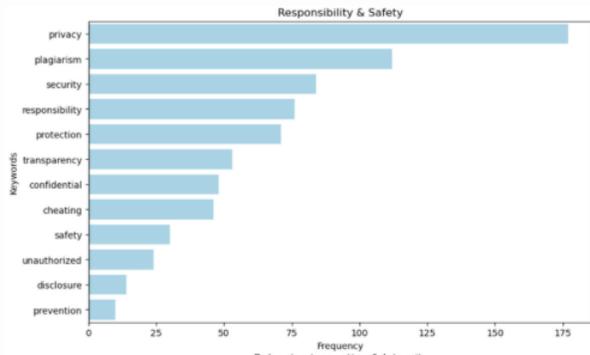
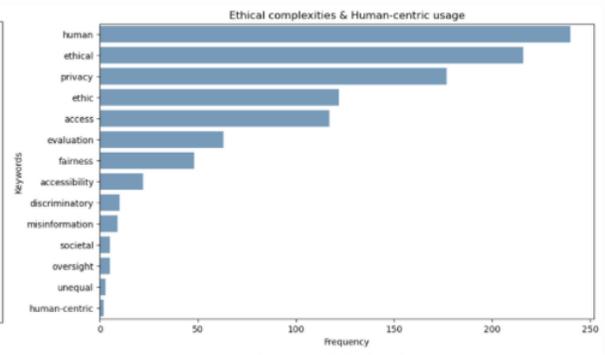
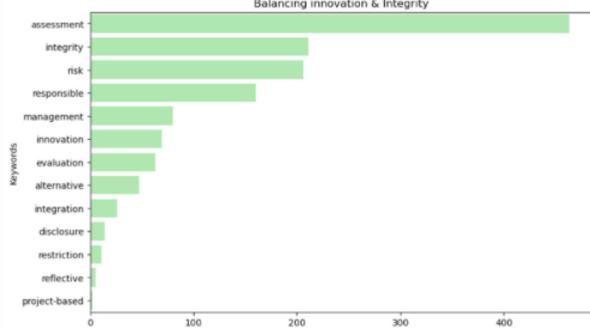
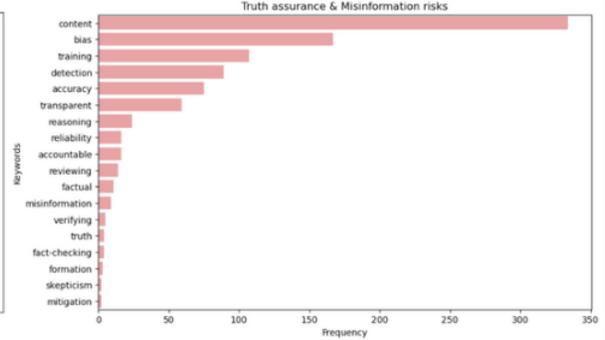
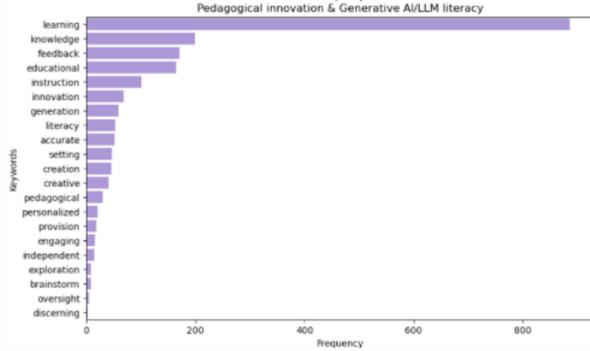
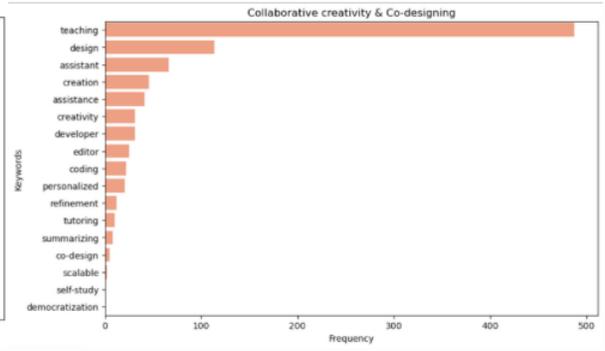
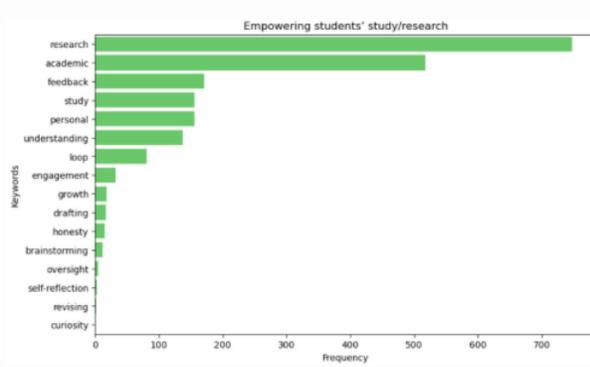
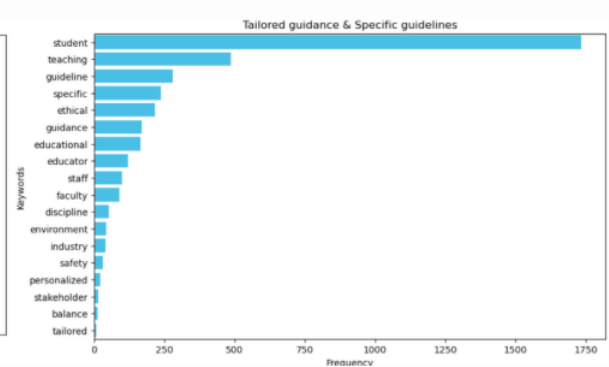

## 4.2. Qualitative Semantic Analysis

After the frequency analysis, we did a qualitative semantic exploration. Compiling impactful concepts from our literature review and discerned the five most crucial concepts for each of the nine categories. Our determination of these key concepts was informed by two criteria: the frequency of their mention in the guidelines, signifying a consensus on their importance (e.g., "integrity," "fairness"), and their unique presence in the discourse, which pointed to their individual significance despite fewer mentions (e.g., "skepticism," "democratization"). Figure 2 illustrates the frequency analysis of these qualitative findings.

**Figure 2**: The Frequency of Key Concepts in the Major Nine Themes

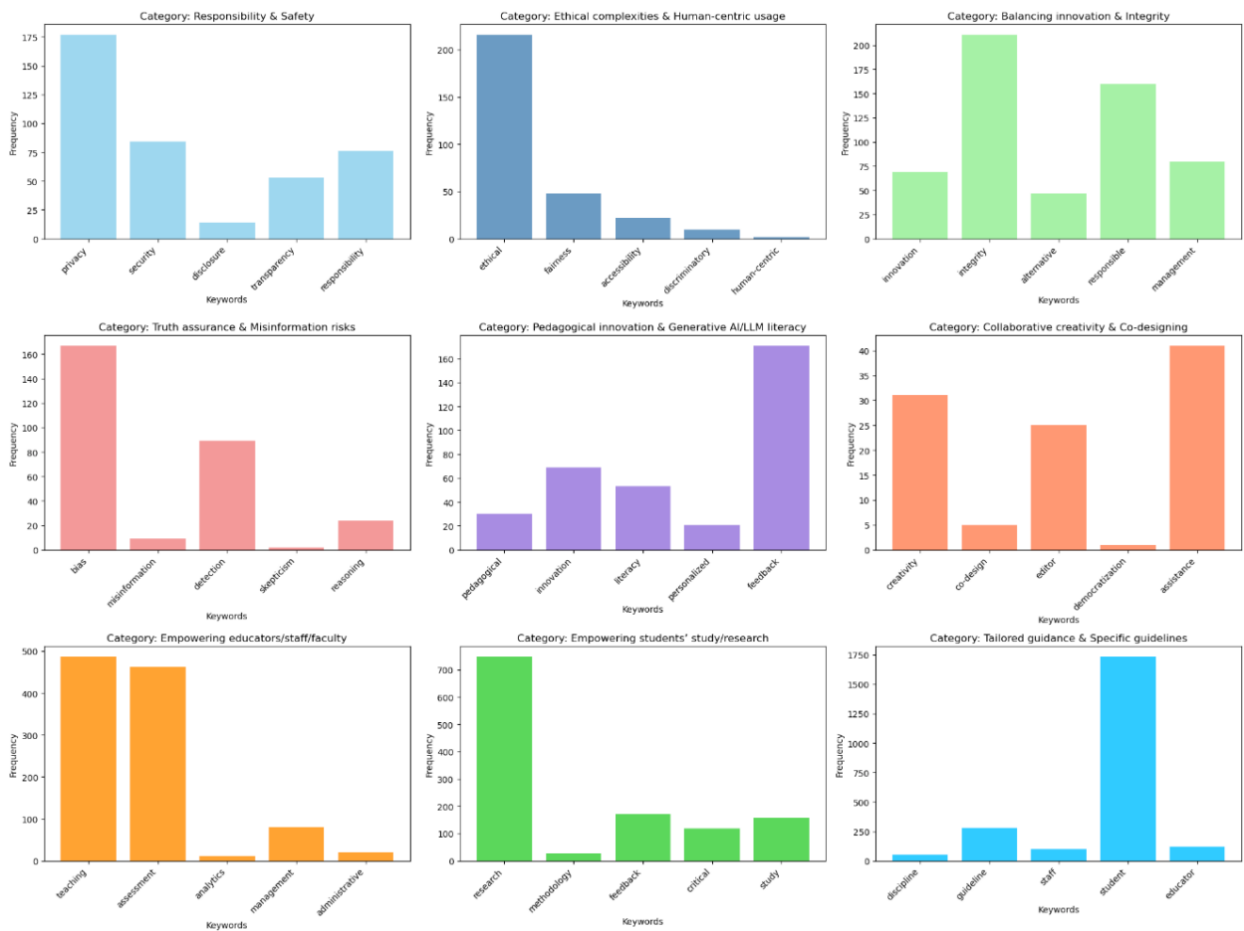

# V. Discussion and Synthesis

In this section, we delve into the multifaceted landscape of integrating GAI and LLM into educational frameworks. Through a synthesis of quantitative and qualitative insights, we highlight key areas requiring attention and development in future guidelines for the effective utilization of GAI and LLMs. Our exploration encompasses crucial themes such as responsibility and safety, ethical complexities, innovation, integrity, truth assurance, collaborative creativity, decision-making paradigms, policy considerations, and the nuanced dynamics of fairness, equality, equity, and access. By rethinking pedagogy, decision-making processes, policies, and navigating the terrain of cautious optimism in AI policies, we aim to strike a balance between tradition and technology while ensuring ethical and effective integration in academia.

## 5.1. Synthesis of Quantitative and Qualitative Insights

Our semantical analysis of the major key concepts, compared with the frequency of keywords in the guidelines text, reveals that despite the low frequency of certain terms, high attention and deeper exploration and development are needed in future guidelines for the effective use of GAI and LLMs.

For instance, the analysis of the 80 GAI/LLM guidelines has shown that "Privacy" is paramount, while "disclosure" ranks lowest in the first category (i.e., responsibility and safety), appearing frequently in 177 and 14 places, respectively, among all the guidelines. While, 'disclosure' would be crucial for responsible GAI/LLM use, emphasizing "disclosure" in the guidelines ensures that users are informed about the capabilities and limitations of AI systems. It also promotes responsible decision-making and accountability among AI developers and users.

The analysis has shown that "ethical" is paramount, while "human-centric" ranks lowest in the second category (i.e., Ethical complexities & Human-centric usage), appearing frequently in 216 and 2 places, respectively, among all the guidelines. Emphasizing 'human-centric' usage and ensuring 'fairness' can secure some major ethical aspects. Prioritizing "human-centric" usage involves incorporating user feedback, designing interfaces for accessibility and inclusivity, and considering ethical concerns such as privacy and consent. Guidelines should also include measures for data fairness, algorithmic transparency, and fairness assessments to promote equitable AI systems across diverse populations.

The examination of the 80 GAI/LLM guidelines has also revealed that "Integrity" stands as paramount, whereas "alternative" methods occupy the lowest position in the third category (i.e., Balancing Innovation & Integrity), appearing frequently in 211 and 47 instances, respectively, across all the guidelines. Exploring 'alternative' methods for teaching, assessment, and research can strike a balance between innovation and integrity. This could include personalized learning pathways, AI-powered grading systems, and innovative research methodologies that leverage AI for data analysis and insights.

Moreover, addressing the low frequent concepts 'misinformation' as well as different aspects and the role of 'skepticism' is vital for truth assurance and risk mitigation. They also are crucial for building trust in AI systems. Guidelines should outline strategies for detecting and combating misinformation, fostering critical thinking skills, and promoting responsible information sharing practices.

The analysis has revealed that "assistance" is of utmost importance, whereas methods for "democratization" rank lowest in the sixth category, Collaborative Creativity & Co-designing. Specifically. In fact, "assistance" appears frequently in 41 instances, while "democratization" methods appear only one time, among all the guidelines analyzed. There is a need to develop more personalized innovation and democratization in GAI/LLM usage, treating users as co-designers. Developing "personalized" innovation involves customizing AI solutions to individual user preferences and needs. This could include personalized recommendations, adaptive learning algorithms, and tailored user experiences that enhance engagement and effectiveness. "Democratization" also involves making these technologies accessible and understandable to a wide range of users. This includes providing educational resources, promoting open-source initiatives, and encouraging collaboration and knowledge-sharing within the AI community.

Predictive analytics can empower staff and educators, enhancing pedagogical management. Leveraging "predictive analytics" empowers staff and educators by providing insights for decision-making, resource allocation, and student support. This can optimize educational outcomes, improve organizational efficiency, and enhance pedagogical management strategies. Moreover, encouraging 'critical' thinking and implementing new AI-powered study methods can empower students. Integrating "critical" thinking skills with AI-powered study methods fosters a deeper understanding of content, encourages analytical reasoning, and promotes creative problem-solving. It enables students to navigate complex information landscapes and make informed decisions.

It is also shown that the 'student' guidelines are the most discussed concepts, while there is much lower discussion about the guidelines for teachers, educators, educational management, etc. Additionally, the discipline-based guidelines have the lowest frequency, indicating a potential significant gap in these guidelines. Discipline-based guidelines tailored to specific fields can further ensure responsible usage of LLMs and GAIs. Customizing guidelines for different fields ensures that AI use matches each field's distinct needs, challenges, and ethical considerations. This includes working with experts in each domain, integrating top industry practices, and adjusting guidelines to progressive technological developments.

## 5.2. Rethinking Pedagogy: Striking a Balance between Tradition and Technology

Educators face what appears to be diametrically paradoxical pedagogical choices: either approach learning with more traditional methods or incorporate AI more often on the assumption that using AI effectively will comprise a vital role in the students' future employment. It is not clear that a class can emphasize the importance of deep thinking without AI while also championing the uses of AI as a helpful resource for thinking--if indeed it is helpful in an empirical sense, rather than merely anecdotal use cases. (See, Table 4 &5)

If the intention of prohibiting the use of LLMs is to force students to grapple with research and forming thoughts, then perhaps such prohibitions make sense. For example, professors may discourage the use of LLMs for academic work because they believe students may become overly dependent on the model outputs, become lazy when conducting research, accept the outputs as sufficiently well-written and comprehensive for a final daft without edits or additions, fail to acquire critical thinking and ideation skills, or suffer some other intellectual atrophy from outsourcing thinking to the model.

But if the intention is to best prepare students to compete in a workforce where their peers may rely heavily on LLMs and understand how to write prompts and confirm the veracity of outputs, then perhaps such prohibitions do a disservice to the students most effectively. It is possible that LLMs will become a normal, frequent, or even pervasive fact of work life, enabling people to perform tasks at superior levels compared to similarly situated workers without access to LLMs. In this scenario, allowing and encouraging students to become proficient or exceptional users of LLMs may position them to excel in the workplace and become more productive members of society without the concomitant decrease in intellectual capacity. They would be learning differently, but not necessarily inferiorly.

**Table 4**: The Education Preparation Paradox: Benefits

| Traditional Approach | GAI-driven Approach |
|---|---|
| Emphasizes deeper thinking | Introduce students to a wider variety of ideas in a shorter amount of time |
| Teaches how to conduct thorough research | May introduce students to a wider variety of ideas in a shorter amount of time |
| Emphasizes the inception, curation, and refining of ideas | Allows students to draft and refine more rapidly, potentially helping them shift focus to the more human-necessary portions of a task |
| Provides students with an opportunity to wrestle with various ideas, understanding how they do or do not relate | Showcases the need to be adaptable to innovative technologies in a constantly changing society |
| Students are more likely to remember information gathered by this method[16] | Prioritize efficiency and productivity |
| Lessens over-reliance on AI for critical thinking/ Advocates a more cautious approach | Embracing Risk: Advocating for Innovative Approaches |
| Follow empirically tested methods | Explore untested but potentially promising methods |
| Rely on the educator to adapt methods to students in relatively small groups | Focus on commodified tasks like brainstorming ideas as opposed to skills like collaborating and critical thinking |

**Table 5**: The Education Preparation Paradox: Drawbacks

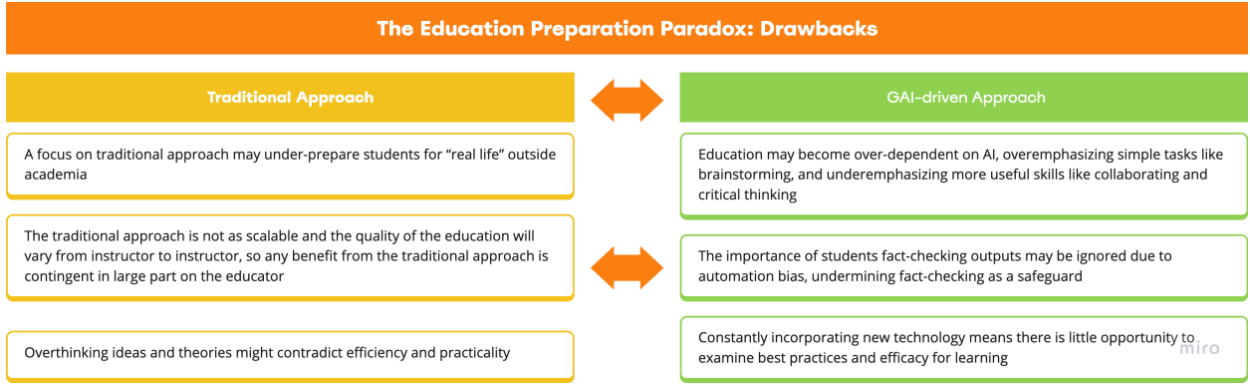

## 5.3. Redefining Decision-Making in Education: Machine-in-the-Loop vs. Human-in-the-Loop

The concept of human-in-the-loop presents another pleasant-sounding but unrevealing preference. It would help to better understand when a human-in-the-loop is desired and compare that with when human-in-the-loop affects outcomes in a positive way, versus a negative outcome or no difference at all. In many instances, we might expect automation bias to lead humans to simply accept whatever output or recommendation the AI produces. This is, after all, the root of the problem with hallucinations. In sum, it is not clear that having a human-in-the-loop will make a meaningful difference in many scenarios.

However, an alternative approach, coined "Evaluative AI," replaces the human-in-the-loop with a machine-in-the-loop, and this approach may be more beneficial to educational institutions. Under the Evaluative AI approach, the AI would provide the user with evidence for and against certain decisions, rather than make a recommendation. This helps the user retain agency while still providing potentially more helpful insights than humans alone would create. It also shifts the responsibility and accountability for decisions more fully onto humans, allowing society a more effective means of reviewing and improving education systems.

In both approaches the timing of the intervention—the point when the AI either makes the recommendation or presents the evidence—can play a vital role.[95] This suggests that blanket support for either approach would be too general. How the chosen approach is implemented may be as important or even more important than which approach is used (See, Table 6).

**Table 6.** Educational Decision-Making: Human-in-the-Loop vs Machine-in-the-Loop approach

| Dimensions | Description | | |
|---|---|---|---|
| | **Autonomous Decision** | **Human-in-the-Loop** | **Machine-in-the-Loop** |
| What it is | Total reliance on AI to make best decisions | AI develops recommendations | AI provides arguments for both sides |
| Benefits | Fastest and possibly simplest approach. Requires the least time and effort of humans. | Ensures a human has an opportunity to review. May also require human sign-off before a decision is final. | Requires humans to contemplate options and that humans make the decisions |
| Drawbacks | AI could be biased, untrustworthy, hacked, or any number of other problems | Possibility for automation bias (over-trusting the AI and simply agreeing with its recommendation each time) | Possibility for false equivalency (presenting both sides as if they are equally valid even when that is not true) |
| Efficacy | Not clearly better than either of the alternative approachesNot clearly better than either of the alternative approaches | Not clearly more useful than machine-in-the-loop in most instances | Not clearly more useful than human-in-the-loop in most instances |

## 5.4. Rethinking Policies: From Uniformity to Diversity

While institutions may want to encourage students to deeply engage with content, it is unclear how they can. Students can just as easily ask an LLM to summarize a long article or research paper as they can ask it to provide a response to a quiz question. It may be that deep discussions of topics, allowing educators to probe student comprehension at more than a surface level, may be the best or only way to ensure students are grappling with the content. However, such discussions are not scalable and may require more time than shorter classes provide. One solution could be to hire more educators and decrease the student to educator ratio, but such an approach may not be financially tenable in most instances.

While policies may necessarily differ between departments, it is not clear why or in what way. It may be worth exploring whether the assumption of different policies stands scrutiny, or whether a more uniform approach (either generally open to AI or generally closed) would be just as effective. Some departments, such as math, may use AI less for the simple reason that LMs are not great at math, but that does not mean a separate policy is needed, just that the use of AI will be self-limiting.

## 5.5. The Complexity of Fairness: Perspectives on Equality, Equity, and Access

A discussion on fairness alone could yield an entire research paper. When institutions declare they are searching for a fair product, they must first decide what they mean by the term. Consider just two of over a dozen ways to define fairness: pursuing equal outcomes (every student treated the same) and equitable outcomes (students who require more resources to learn are provided with more resources) would both be examples of fairness and yet equality and equity can also be impossible to reconcile. Unless an institution knows what it is looking for, claims of seeking fairness, privacy, accessibility, and more do not reveal much in the way of substance.

It would be interesting to know what bridging the digital divide entails. Most American children have a smartphone by age 11, and that percentage shoots up sharply to over 90% as they enter high school.[96], [97] Smartphones, in turn, are capable of accessing the internet, including the most prominent LLMs and learning tools used by schools. Access to AI has never been more widespread. While there are no doubt millions of children at a technological disadvantage, this may also provide the means for a natural experiment to compare the AI haves vs have-nots and see if AI improves educational and life outcomes. Given that so many educators still prohibit the use of AI, one hypothesis may be that the digital divide actually *benefits* students with less access to technology.[98]

## 5.6. Navigating Cautious Optimism: AI Policies in Academia

The overall theme of these policies seems to be one of cautious optimism. Despite acknowledging the potential for harms in the form of privacy, fairness, bias, and more, academia remains open to students using the LLMs in many cases, with more than two dozen seeing potential for AI to improve education through personalization and others seeing such varied possibilities as improving research and brainstorming ideas. Given that virtually no institution bans the use of AI while all of them suggest ways to use it wisely, it suggests their fears of potential harm, though real, may be overstated.

Some advocated uses, such as with grading, seem to be unfounded at the moment. While there is certainly the potential for AI to improve grading and to provide more timely feedback, thereby leading to better student outcomes, this theory has yet to be proven with empirical research. It again shows that institutions may be less cautious about the potential harms they warn against. In areas such as grading, it may be prudent for institutions to establish policies that limit such use to low-stakes assignments intended to assess students' formative knowledge, as opposed to grading assignments or final exams that could have significant impacts on final grades.

# VI. Concluding Remarks and Future Directions

The incorporation of GAI and LLM into academic education brings both opportunities and challenges. Our research emphasizes the importance of balanced strategies that leverage the advantages of these technologies while addressing ethical issues, ensuring fair access, and enhancing educational outcomes. As we explore this area, several key themes emerge that deserve more investigation and advancement in future research and policies.

Firstly, discussions about integrating GAI and LLM should prioritize responsible innovation and ethical practices. This means that we should consistently strive for transparency, accountability, and users, especially concerning privacy, disclosure, and human-centric use. It is crucial to continue our research into ways to ensure fairness, equity, and access for all groups to prevent biases and create inclusive educational settings.

Secondly, redefining decision-making in education, whether involving machines or humans in the loop, needs careful consideration. Future research should thoroughly assess the effectiveness and impacts of these decision-making methods, investigating their influence on educational outcomes, student involvement, and the development of critical thinking skills.

Thirdly, policies regarding GAI and LLMs in academic education should shift from one-size-fits-all to customized approaches, reflecting the specific requirements of various fields and educational contexts. This includes reconsidering how AI is used in teaching, assessment, and research, and finding new methods that use AI while preserving academic integrity and fostering meaningful learning experiences.

Finally, adopting cautious optimism in AI policies demands continuous dialogue, teamwork, and empirical research. Institutions need to find a middle ground between leverage AI's educational benefits and mitigating possible risks. Future efforts should prioritize empirical studies to evaluate how integrating AI affects student learning, teaching practices, and institutional policies.

Looking ahead, working together across disciplines, involving stakeholders, and relying on evidence-based approaches will be essential for responsibly integrating GAI and LLMs into education. This paper serves as a starting point for these discussions, urging scholars, policymakers, educators, and technologists to keep exploring these complex issues and guiding ethical and inclusive AI integration in academia.

**Conflict of interest**: The authors declare that the research was conducted in the absence of any commercial or financial relationships that could be construed as a potential conflict of interest.

**Acknowledgements**
This research is funded by the National Science Foundation (NSF) under grant number 2125858. The authors express their gratitude for the NSF's support, which made this study possible. Furthermore, in accordance with MLA (Modern Language Association) guidelines, we note the use AI-powered tools, such as OpenAI's applications, for assistance in editing and brainstorming.

**Institutional Review Board Statement:** Not applicable.
**Informed Consent Statement:** Not applicable.
**Data Availability Statement:** Not applicable.

[35] U. of Pennsylvania, "Statement on Guidance for the University of Pennsylvania Community on Use of Generative Artificial Intelligence | UPenn ISC." 2023. [Online]. Available: https://www.isc.upenn.edu/security/AI-guidance

[36] L. A. University of California, "Guidance for the Use of Generative AI – UCLA Center for the Advancement of Teaching." 2023. [Online]. Available: https://teaching.ucla.edu/resources/ai_guidance/

[37] C. University, "Artificial Intelligence (AI) | IT@Cornell." 2023. [Online]. Available: https://it.cornell.edu/ai

[38] U. of Toronto, "Generative Artificial Intelligence in the Classroom: FAQ's – Office of the Vice-Provost, Innovations in Undergraduate Education." 2024. [Online]. Available: https://www.viceprovostundergrad.utoronto.ca/16072-2/teaching-initiatives/generative-artificial-intelligence/

[39] U. of B. Columbia, "UBC Guidance - Generative AI." 2023. [Online]. Available: https://genai.ubc.ca/guidance/

[40] M. University, "Principles on Generative AI in Teaching and Learning at McGill." 2023. [Online]. Available: https://www.mcgill.ca/provost/files/provost/principles_on_generative_ai_in_teaching_and_learning_at_mcgill.pdf

[41] U. of Alberta, "AI-Squared - Artificial Intelligence and Academic Integrity | Centre for Teaching and Learning." 2023. [Online]. Available: https://www.ualberta.ca/centre-for-teaching-and-learning/teaching-toolkit/teaching-in-the-context-of-ai/artificial-intelligence-academic-integrity.html

[42] U. of Waterloo, "Artificial intelligence and ChatGPT | Academic Integrity." 2023. [Online]. Available: https://uwaterloo.ca/academic-integrity/artificial-intelligence-and-chatgpt

[43] U. of Montreal, "About - Déclaration de Montréal IA responsable." 2017. [Online]. Available: https://montrealdeclaration-responsibleai.com/about/

[44] M. University, "Provisional Guidelines on the Use of Generative AI in Teaching and Learning." 2023. [Online]. Available: https://provost.mcmaster.ca/office-of-the-provost-2/generative-artificial-intelligence/task-force-on-generative-ai-in-teaching-and-learning/provisional-guidelines-on-the-use-of-generative-ai-in-teaching-and-learning/

[45] U. S. D. of Education, "Artificial Intelligence and the Future of Teaching and Learning." 2023. [Online]. Available: https://www2.ed.gov/documents/ai-report/ai-report.pdf

[46] U. del Rosario, "Guidelines for the Use of Artificial Intelligence in University Courses." 2023. [Online]. Available: https://forogpp.files.wordpress.com/2023/02/guidelines-for-the-use-of-artificial-intelligence-in-university-courses-v4.3.pdf

[47] U. of B. Aires, "Guidelines for the use of ChatGPT and text generative AI in Justice." 2023. [Online]. Available: https://ialab.com.ar/wp-content/uploads/2024/02/Guia-uso-IAG-.pdf

[48] U. de L. Andes, "Guidelines for the use of artificial intelligence in university contexts." 2023. [Online]. Available: https://juangutierrezco.files.wordpress.com/2023/08/guidelines-for-the-use-of-artificial-intelligence-in-university-contexts-v5.0.pdf

[49] P. J. University, "Editorial Pontificia Universidad Javeriana. Editorial Policy, Publication Ethics and Malpractice Statement." 2023. [Online]. Available: https://revistas.javeriana.edu.co/files-articulos/CRC-EPUJ/manuales/ETHICS/

[50] U. de san andres, "COMPILED: READINESS OF THE JUDICIAL SECTOR FOR ARTIFICIAL INTELLIGENCE IN LATIN AMERICA." 2023. [Online]. Available: https://cetys.lat/wp-content/uploads/2021/09/compilado-eng.pdf

[51] G. and D. T. Secretariat, "NATIONAL ARTIFICIAL INTELLIGENCE STRATEGY." 2021. [Online]. Available: https://cdn.www.gob.pe/uploads/document/file/1909267/National%20Artificial%20Intelligence%20Strategy%20-%20Peru.pdf

[52] T. Ministry of Science, "Guidelines for the use of artificial intelligence tools in the public sector." 2023. [Online]. Available: https://minciencia.gob.cl/uploads/filer_public/ae/9a/ae9a7ce7-807b-4781-9ac3-9b253bfbe735/of_n711_2023_dis_lin_ia_minciencia.pdf

[53] pontificia universidad católica de chile, "ChatGPT: How to use it in classes?" 2023. [Online]. Available: https://www.uc.cl/noticias/chatgpt-como-usarlo-en-clases/

[54] T. University, "About the Forum-INSTITUTE FOR AI INTERNATIONAL GOVERNANCE TSINGHUA UNIVERSITY." 2022. [Online]. Available: https://aiig.tsinghua.edu.cn/en/International_Forum/2022/About_the_Forum.htm

[55] N. U. of Singapore, "B_Oct_Responsible_Use_of_AI - Centre for Technology, Robotics, Artificial Intelligence and the Law." 2023. [Online]. Available: https://law.nus.edu.sg/trail/responsible-use-of-ai/